\documentclass{aa}
\usepackage{graphicx}
\usepackage{natbib}
\usepackage{txfonts}
\begin{document}
   \title{Chemical abundances of the PRGs UGC7576 and UGC9796.}

   \subtitle{I. Testing the formation scenario}

   \author{M. Spavone
          \inst{1,2}\fnmsep\thanks{e-mail: spavone@na.infn.it}
          \and
          E. Iodice\inst{2}
          \and M. Arnaboldi\inst{3,4}
          \and G. Longo\inst{1,2}
          \and O. Gerhard\inst{5}
          }

   \institute{Dipartimento di Scienze Fisiche,
   Universit\'a Federico II, via Cinthia 6, I-80126 Napoli, Italy\\
         \and
             INAF-Astronomical Observatory of Naples, via Moiariello 16,
   I-80131 Napoli, Italy\\
             \and
             European Southern Observatory,
   Karl-Schwarzschild-Stra$\ss$e 2, D-85748 Garching bei M\"{u}nchen,
   Germany\\
   \and INAF, Osservatorio Astronomico di Pino Torinese,
   I-10025 Pino Torinese, Italy\\
   \and Max-Plank-Institut f\"{u}r
   Extraterrestrische Physik, Giessenbachstra$\ss$e, D-85741 Garching
   bei M\"{u}nchen, Germany\\
             }

   \date{Received February, 2011; accepted April, 2011}


 \abstract
{The study of the chemical
 abundances of HII regions in polar ring galaxies and their
 implications for the evolutionary scenario of these systems has been
 a step forward both in tracing the formation history of the galaxy
 and giving hints on the mechanisms at work during the building of
 disk by cold accretion process. It's now important to establish
 whether such results are typical for the class of polar disk galaxies
 as whole.}
{The present work aims at checking the cold
   accretion of gas through a ``cosmic filament'' as a possible
   scenario for the formation of the polar structures in UGC7576 and
   UGC9796. If these form by cold accretion, we expect the HII regions
   abundances and metallicities to be lower than those of
   same-luminosity spiral disks, with values of the order of $Z \sim
   1/10\ Z_{\odot}$, as predicted by cosmological simulations.}
{We have used deep long-slit spectra,
     obtained with DOLORES@TNG in the optical wavelengths, of the
     brightest HII regions associated with the polar structures to
     derive their chemical abundances and star formation rate. We used
     the \emph{Empirical methods}, based on the intensities of easily
     observable lines, to derive the oxygen abundance $12+log(O/H)$ of
     both galaxies. Such values are compared with those typical for
     different morphological galaxy types of comparable luminosity.}
{The average metallicity values for
UGC7576 and UGC9796 are $Z = 0.4 Z_{\odot}$ and $Z = 0.1 Z_{\odot}$
respectively. Both values are lower than those measured for ordinary
spirals of similar luminosity and UGC7576 presents no metallicity
gradient along the polar structure. These data, toghether with other
observed features, available for the two PRGs in previous works, are
compared with the predictions of simulations of tidal accretion,
cold accretion and merging, to disentangle between these scenarios.}
   {}


   \maketitle
%

\section{Introduction} \label{intro}

How galaxies acquire their gas is an open issue in the models of
galaxy formation: recent theoretical works, supported by many
numerical simulations, have argued that cold accretion plays a major
role (\citealt{Kat93}; \citealt{Kat94}; \citealt{Ker05};
\citealt{Dek06}; \citealt{Dek08}; \citealt{Bou09}). \citealt{Ker05}
studied in detail the physics of the {\it cold mode} of gas accretion
and they found that it is generally directed along filaments, allowing
galaxies to draw gas from large distances. In particular, the cold
accretion is a key mechanism to provide gas to disk galaxies
\citep{Bro09}.

Recent simulations of disk formation in a cosmological context
performed by \cite{Age09} revealed that the so called chain-galaxies
and clump-clusters, found only at higher redshifts \citep{Elm07}, are
a natural outcome of early epoch enhanced gas accretion from cold
dense streams as well as tidally and ram-pressured stripped material
from minor mergers and satellites. This freshly accreted cold gas
settles into a large disk-like systems. This scenario reproduces the
observed morphology and global rotation of disks, predicts a realistic
metallicity gradient, and a star formation rate (SFR) of $20
M_{\odot}/yr$. \cite{Age09} found solar metallicity for the inner
disk, while that in the clump forming region is only $\sim 1/10
Z_{\odot}$ due to the accretion of pristine gas in the cold streams
mixed with stripped satellite gas.

Simulations also show that the interaction region between the
new-formed disk and the cold streams can also result misaligned with
the initial galactic disk: based on a very limited statistics,
\cite{Age09} suggest that this misalignment might not be typical, and
it is due to a third cold stream that is perpendicular to the main
filament. More recent analysis show that the accretion of gas along
misaligned filaments with respect to the disk plane are more common
and it leaves traces down to low redshift (\citealt{Dek09N};
\citealt{Ros10}). An almost polar ring can result just as an extreme
case of such process and, as suggested by \cite{Age09}, it could be
responsible for the formation of polar disks.

Hydrodynamical
simulations performed by \cite{Mac06} and \cite{Bro08} have shown that
the formation of a polar disk galaxy can occur naturally in a
hierarchical universe, where most low-mass galaxies are assembled
through the accretion of cold gas infalling along a filamentary
structures. According to \cite{Mac06}, the polar disk forms from cold
gas that flows along the extended $\sim 1 Mpc$ filament into the
virialized dark matter halo. The gas streams into the center of the
halo on an orbit that is offset from radial infall. As it reaches the
center, it impacts with gas in the halo of the host galaxy and with
the warm gas flowing along the opposite filament. Only the gas
accreted perpendicular to the major axis of the potential can survive
for more than a few dynamical lifetimes.

\cite{Bro08} argued that
polar disk galaxies are extreme examples of the misalignment of
angular momentum that occurs during the hierarchical structure
formation: an inner disk starts forming shortly after the last major
merger at $z \sim 2$. Due to its gas rich nature, the galaxy rapidly
forms a new disk whose angular momentum is determined by the merger
orbital parameters. Later, gas continues to be accreted but in a plane
that is almost perpendicular to the inner disk. At $z \sim 0.8$ the
central galaxy is still forming stars in a disk, while the bulk of new
star formation is in a highly inclined polar disk. By $z\sim 0.5$
the inner disk has exhausted its gas, while gas continues to fall onto
the polar disk. From this point on, star formation occurs exclusively in
the polar disk, that remains stable for at least 3 Gyrs. The
formation mechanisms described above can self-consistently explain both
morphology and kinematics of a polar disk galaxy. In particular, the
predictions turn out to be consistent with many features (like colors
and color gradients, longevity, spiral arms, HI content and
distribution) observed for the polar disk galaxy NGC4650A.

NGC4650A is the prototype of the wide Polar Ring Galaxies (PRGs) and has
been studied in detail to understand its formation and physical
properties (\citealt{Arn97}; \citealt{Gal02}; \citealt{Iod02};
\citealt{Swa03}; \citealt{Bou03}; \citealt{Iod06}) in order to
constrain many of the processes at work during galaxy interactions and
merging.  Very recently, \cite{Spav10}, by using deep longslit
spectroscopy with FORS2@ESO-VLT, studied the abundance ratios and
metallicities of the HII regions associated to the polar disk in
NGC4650A, in order to test the cold accretion scenario for this
object. The chemical abundance is one of the key parameters that can
be estimated in a galaxy disk and directly compared with the
theoretical predictions: in fact, if the gas is essentially primordial
it should have a very low abundance of heavy elements.

Main results obtained for NGC4650A show i) that it has metallicity
lower than spiral galaxy disks of the same total luminosity, where $Z
= 0.2Z_{\odot}$, which is consistent with values predicted for the
formation of disks by cold accretion processes \citep{Age09}; ii) the
absence of any metallicity gradient along the polar disk, which
suggests that the metal enrichment is not influenced by the stellar
evolution of the older central spheroid and, thus, the disk was formed
later by the infall of metal-poor gas from outside which is still
forming the disk. The latter is also a characteristic found in some
other PRGs \citep{Bros09} and in LSB galaxies \citep{deB98}, that
  have colors, metallicities, age and brightness similar to those of
  PRGs suggesting that the infall of metal-poor gas may reasonably fit
  all these observational evidences. \citet{Spav10} lead the way to
implement a test for the cold accretion. Such study has revealed an
indirect and well-based check for the cold accretion scenario of disk
formation. In this work we would like to perform the same kind of
analysis on the two PRGs UGC7576 and UGC9796. These two objects are
similar to NGC4650A; such a morphological similarity may suggest a
similar formation mechanism: the polar structures are more extended
with respect to the central spheroids, thus, as NGC4650A, they were
classified as wide PRG \citep{Whi90}; they are characterized by
exponential surface brightness profiles, very blue colors, knotty
appearance and large amount of HI gas all associated to these
components. The most relevant feature in UGC9796 is that the HI
distribution resemble that of a disk in differential rotation, rather
than a ring (\citealt{Res94p}; \citealt{Cox06}), like in NGC4650A
\citep{Arn97}. Given the new results obtained for NGC4650A it is now
important to establish whether the lack of metallicity gradient and
low metallicity, and the inferred accretion of low-metallicity
material after the formation of the central spheroid, are typical for
the class of wide PRGs as a whole: to this aim we have studied the
PRGs UGC7576 and UGC9796, that are suitable systems to investigate
this question further.

\subsection{UGC7576} \label{7576}

UGC7576 (Fig. \ref{7576}) is a kinematically confirmed Polar Ring
Galaxy \citep{Whi90} and its main properties are listed in
Tab. \ref{global}. This object is at a distance of about 94 Mpc, based
on $H_{0} = 75 \ km \ s^{-1} \ Mpc^{-1}$ and heliocentric radial
velocity of $V = 7022\ km \ s^{-1}$, which implies that 1 arcsec =
0.45 kpc. The kinematics shows that the polar structure of
UGC7576 is more similar to a ring rather than a disk, given that
  the rotation curve does not show differential rotation, contrary to
those of NGC4650A and UGC9796. \cite{Res94p} studied the global
morphology of UGC7576 and they were able to distinguish three
components: \emph{i}) the main central body with elliptical isophotes,
\emph{ii}) a narrow polar ring crossing the central region of the main
galaxy, \emph{iii}) a faint outer envelope. The gradient of the
surface brightness distribution of the central galaxy decrease at $r
\geq\ 20''$, while both B-V and V-R color indices slightly decrease
with radius.

The distribution of color
indices is slightly asymmetric, suggesting that the polar ring is
projected not exactly on the nucleus but with a small NW
displacement. The color asymmetry correlates with the asymmetry in the
hydrogen distribution \citep{Sch84}. Moreover, the plane of the disk
reveals a warping, which suggests that the ring has not settled yet in
the equilibrium plane.

Surface brightness distribution along the major axis of the ring is
very symmetric and the surface brightness is approximately constant
($\mu_{B} \sim\ 24.3$) up to $r \sim\ 35''$, then falls abruptly. Also
the colors have no evident changes within the region of constant brightness.

UGC7576 is also embedded in a faint outer
envelope, whose major axis position angle coincides with the position
angle of the polar ring and this suggests a common origin for both
components \citep{Res94p}.

\cite{Res94} have analyzed the $H_{\alpha}$ spectroscopy for a
sample of PRGs, including both UGC7576 and UGC9796.  In the case of
UGC7576, the $H_{\alpha}$ rotation curve along the polar structure was
consistent with the HI rotation curve obtained by \cite{Sch84} and
both curves are straight lines with the same gradient: this let the
authors to conclude that the ring is, more reliable, an edge-on
narrow annulus.  By assuming a spherical mass distribution,
\cite{Res94} estimated a total mass-to-light ratio of $M/L_{B} \sim
11.3 M_{\odot}/L_{\odot}$ within the last measured point of the
rotation curve (where the velocity reaches its maximum value): this
value, together with the $M/L_{B}$ for UGC9796, is
among the highest $M/L_{B}$ estimated for the whole sample of PRGs.
\cite{Res94} also performed a simple dynamical mass model for both PRGs
studied in this work: in the hypothesis of a spherical dark halo, for
UGC7576 they found a dark-to-luminous mass ratio of about 1.3, inside
the polar ring radius.

\begin{figure*}
\centering
\includegraphics[width=10cm]{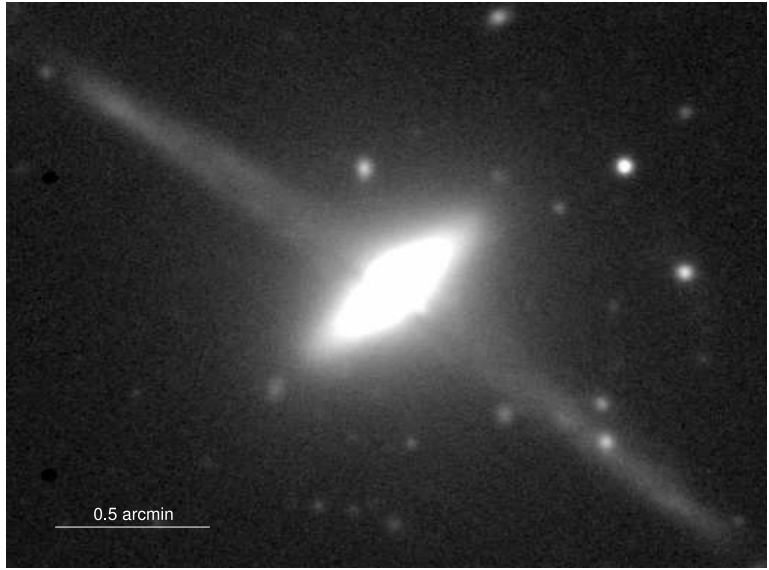}
\caption{R-band image of the Polar Ring Galaxy UGC7576} \label{7576}
\end{figure*}

\subsection{UGC9796} \label{9796}

UGC9796 (Fig. \ref{9796}), also known as IIZw 73, is at a distance of
about 72 Mpc based on $H_{0} = 75 \ km \ s^{-1} \ Mpc^{-1}$ and
heliocentric radial velocity of $V = 5406\ km \ s^{-1}$, which implies
that 1 arcsec = 0.35 kpc. It has one of the most non-polar PRG. Its
apparent major axis in fact is only $65^{\circ}$ from the major axis of
the central S0 rotationally supported galaxy and this implies a rapid
rate of differential precession. The polar structure is less symmetric
than in UGC7576 and the distribution of the colors is also asymmetric,
with the NE side considerably redder than the SW side. The color
asymmetry coincides with the asymmetry of the HI density
distribution. The HI gas is all associated with the polar structure,
which thus contains as many baryons as the host galaxy \citep{Sch84},
and shows a central hole at about $25''$. The huge mass-to-light ratio
of $M_{dyn}/L_{B} \simeq 50$ in solar units, lead \cite{Cox06} to
conclude that most of the mass in this system is dark. As in the case
of UGC7576, the $H_{\alpha}$ rotation curve \citep{Res94} is in good
agreement with the HI rotation curve \citep{Sch84}. The shape of the
rotation curve indicates that this component is actually a
differentially rotating disk, rather than a ring, very similar to the
polar disk in NGC4650A.

In Tab. \ref{global} we summarize the photometric and HI observed quantities for UGC7576 and
UGC9796 and we add those of NGC4650A for reference in the same table.

\begin{figure*}
\centering
\includegraphics[width=8cm]{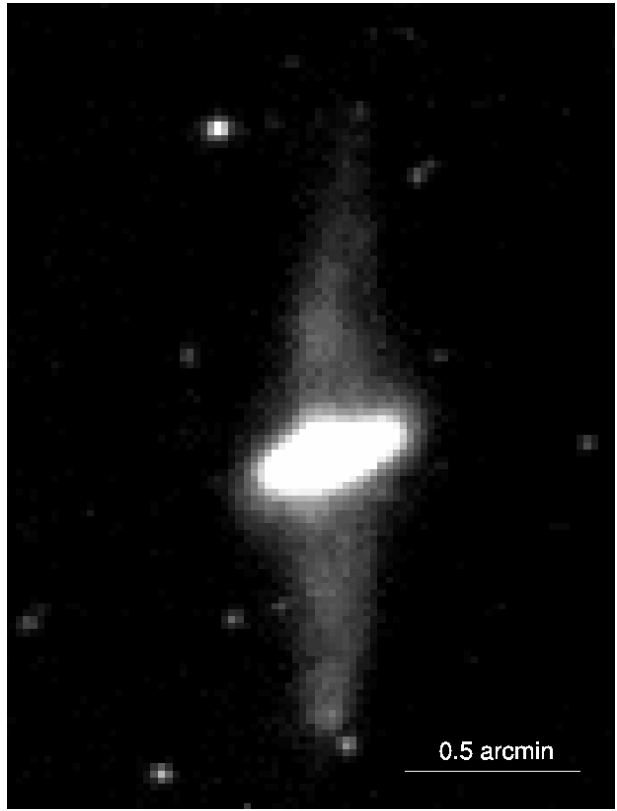}
\caption{Optical image of the Polar Ring Galaxy UGC9796, obtained by the authors at TNG telescope.} \label{9796}
\end{figure*}

\begin{table*}[h]
\caption{\label{global}Global properties of UGC7576 and UGC9796
compared to those observed for NGC4650A.} \centering
\begin{tabular}{lcccc}
\hline\hline
Parameter&UGC7576&UGC9796& \vline & {\it NGC4650A}\\
\hline
R.A. (J2000)           &12h27m41.8s    &15h15m56.3s & \vline & {\it 12h44m49.0s}\\
Decl. (J2000)           &+28d41m53s   &+43d10m00s & \vline &{\it -40d42m52s}\\
Helio. radial velocity &7022 km/s     &5406 km/s & \vline & {\it 2880 km/s}\\
Redshift         &0.02342 &0.01832   & \vline & {\it 0.009607}\\
Distance      &94 Mpc     &72 Mpc& \vline & {\it 38 Mpc}\\
\emph{Central galaxy}&&&\vline&\\
$M_{B}$ & -19.15& -17.93 & \vline & {\it -18.83}\\
B-V & +0.84 & +0.92 & \vline & {\it +0.78}\\
V-R & +0.46 & +0.55 &\vline &\\
\emph{Polar structure}&&&\vline&\\
$M_{B}$      &-17.5   &-17.0 & \vline & {\it -17.0}\\
$M(HI)(M_{\odot})$  & $2.7 \times 10^{9}$   & $2.6 \times 10^{9}$&\vline & {\it $8.0 \times 10^{9}$}\tablefootmark{a}\\
$M(HI)/L_{B}$ & 0.6 & 1.5 & \vline & {\it 4}\\
B-V & +0.70\tablefootmark{b} & +0.57\tablefootmark{b} & \vline & {\it +0.26}\tablefootmark{c}\\
$\mu_{B}$ & 24.3 & 24.5 & \vline & {\it 22.6}\\
$R_{25}$ & 13.6 & 10.8 & \vline & \it {7.0} \\
$r_{max}$ & $40''$ & $60''$ & \vline & {\it $40''$} \\
\hline
\end{tabular}
\tablefoot{ \tablefoottext{a}{Arnaboldi et al. 1997} \tablefoottext{b}
  {Reshetnikov et al. 1994} \tablefoottext{c} {Iodice et al. 2002}}
\end{table*}

\section{Observation and data reduction} \label{obs}

\begin{figure*}
\includegraphics[width=8cm]{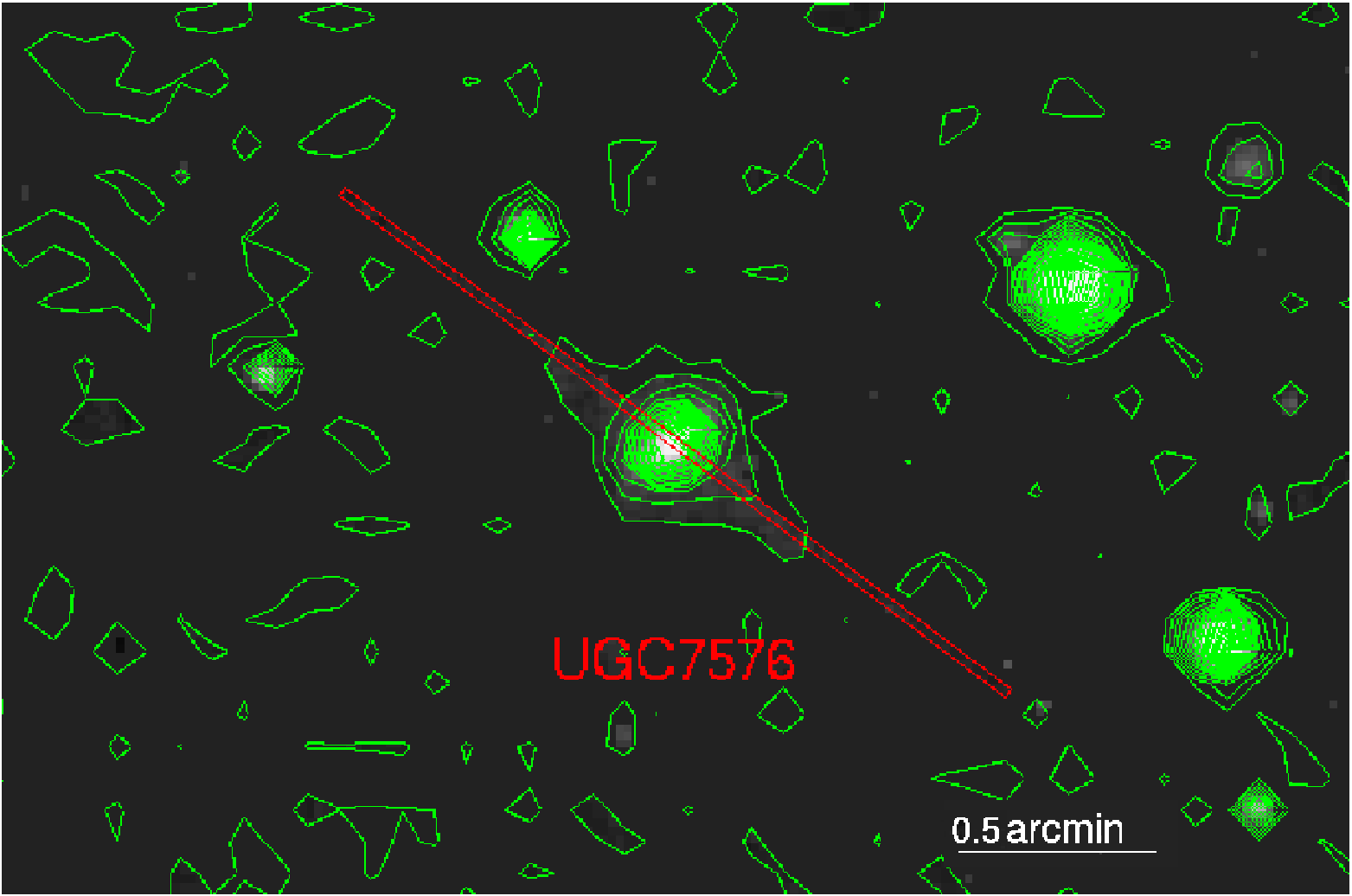}
\includegraphics[width=8cm]{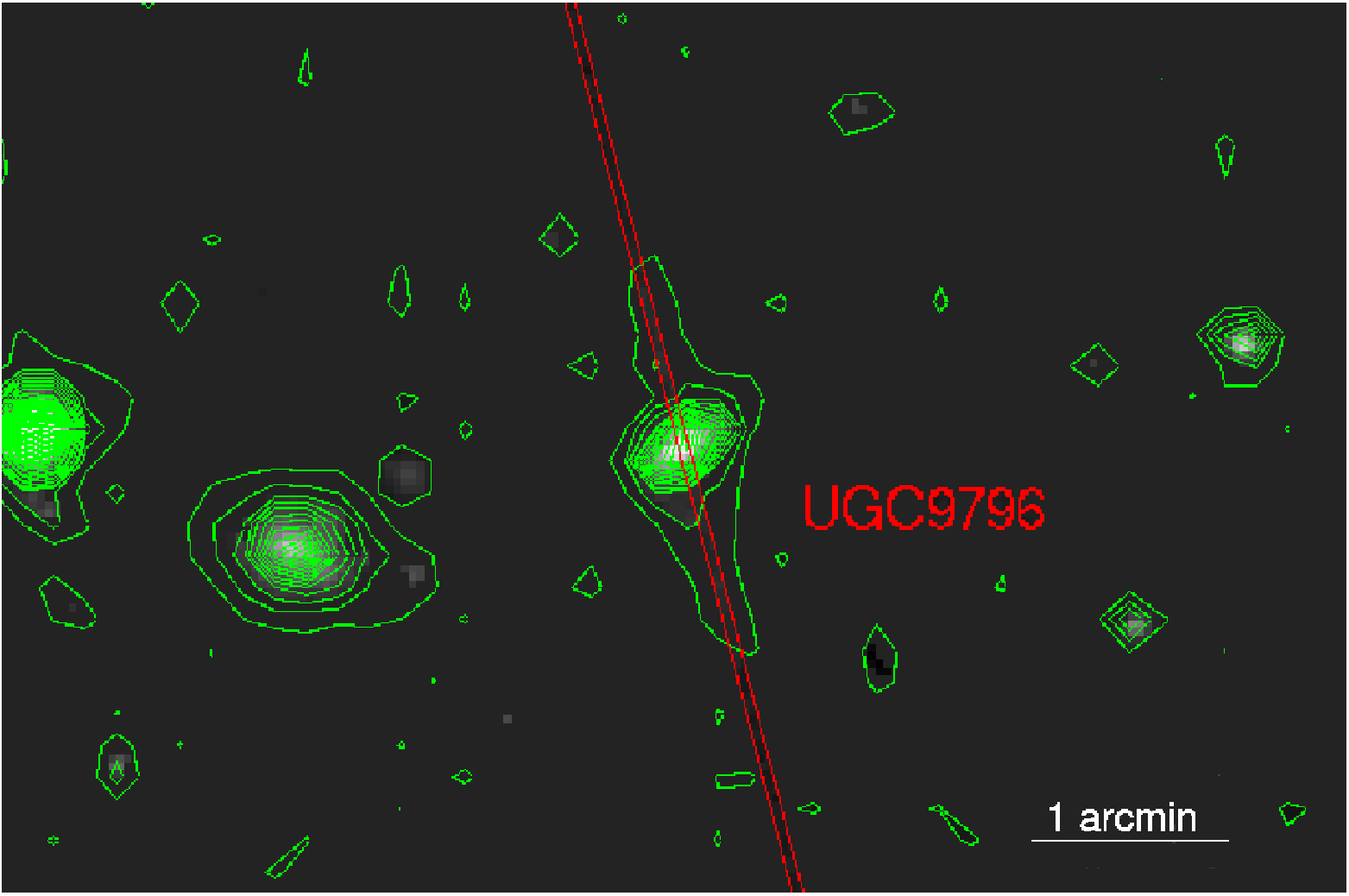}
\caption{Optical image of UGC7576 (left) and UGC9796 (right), with
  superimposed slits used to acquire data analyzed in this
  work.} \label{slit}
\end{figure*}

The spectra analyzed in this work were obtained with DOLORES@TNG
(Device Optimized for the LOw RESolution), in visitor mode during the
observing run A21TAC-54 (on May 2010). DOLORES is installed at the
Nasmyth B focus of the TNG and is equipped with the E2V 4240 CCD with
an angular resolution of $0''.252\ pix^{-1}$. \\The adopted slit was
$2''$ wide and it was aligned along both the North and South side
  of the polar structures of the two galaxies UGC7576 and UGC9796, at
$P.A. = 53^{\circ}$ and $P.A. = 16^{\circ}$ respectively (see
Fig. \ref{slit}), in order to include the most luminous HII regions in
the polar structures. The total integration time for each object is 4 hrs,
with an average seeing of $1''.2$.

At the systemic velocities of UGC7576 and UGC9796, to cover the
red-shifted emission lines of $[OII]\lambda3727$,
$H_{\gamma}(\lambda4340)$, $[OIII]\lambda4363$,
$[OIII]\lambda\lambda4959,5007$, $H_{\beta}(\lambda4861)$ and
$H_{\alpha}(\lambda6563)$, the grism LR-B was used in the
$3600-6800$\AA\ wavelength range, with a dispersion of 2.52 \AA/pix.

The data reduction was carried out using the {\small CCDRED} package
in the IRAF\footnote{IRAF is distributed by the National Optical
  Astronomy Observatories, which is operated by the Associated
  Universities for Research in Astronomy, Inc. under cooperative
  agreement with the National Science Foundation.} ({\it Image
  Reduction and Analysis Facility}) environment. The main strategy
adopted for each data-set included dark subtraction\footnote{Bias
  frame is included in the Dark frame.}, flat-fielding correction, sky
subtraction and rejection of bad pixels. Wavelength calibration was
achieved by means of comparison spectra of Hg+Ne lamps acquired for
each observing night, using the IRAF TWODSPEC.LONGSLIT package. The
sky spectrum was extracted at the outer edges of the slit, for $r \ge
30$ arcsec from the galaxy center, where the surface brightness is
fainter than $24 mag/arcsec^2$, and subtracted off each row of the two
dimensional spectra by using the IRAF task BACKGROUND in the
TWODSPEC.LONGSLIT package.  On average, a sky subtraction better than
$1\%$ was achieved. The sky-subtracted frames were co-added to a final
median averaged 2D spectrum.

\begin{figure*}
\includegraphics[width=8cm]{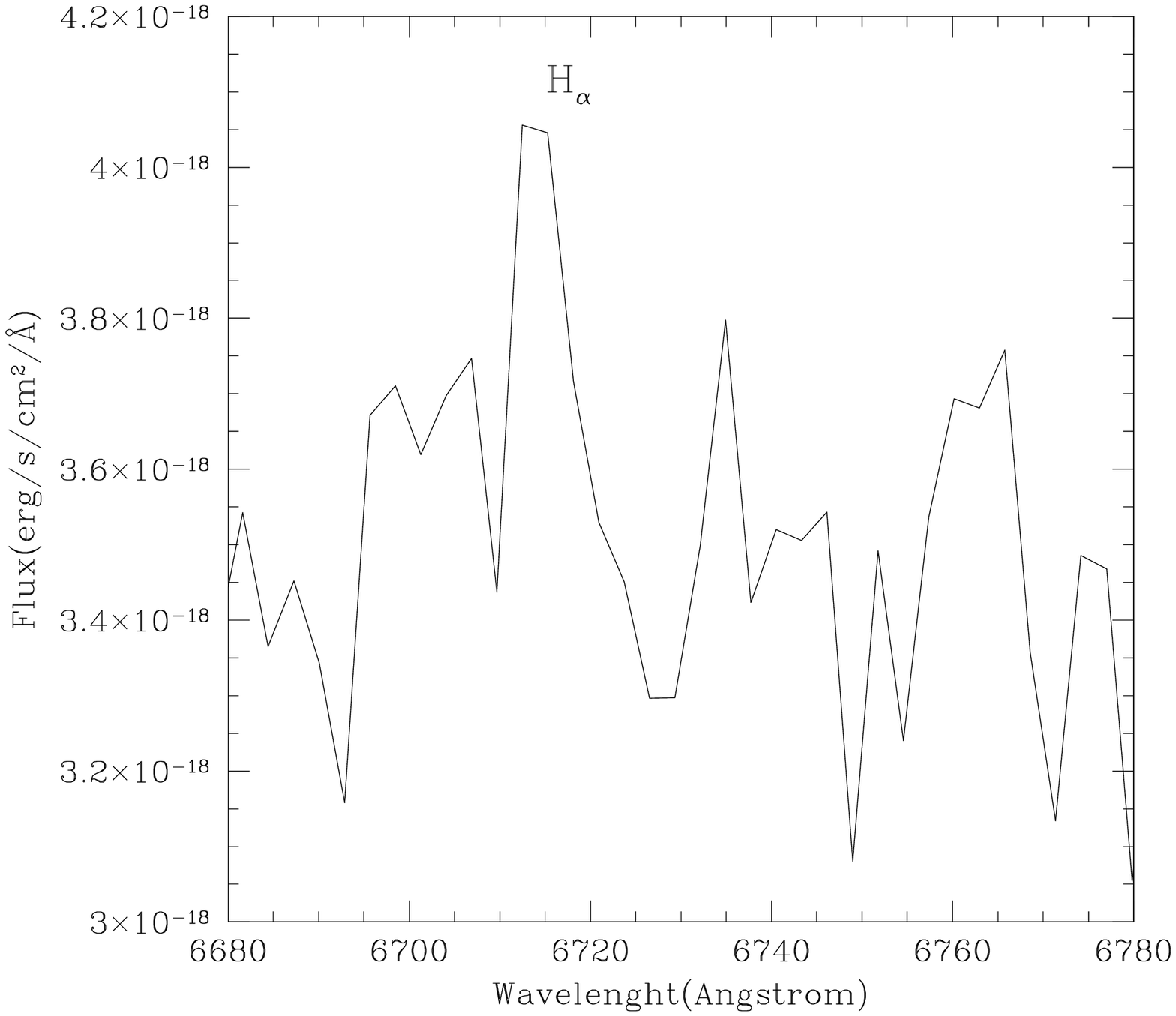}
\includegraphics[width=8cm]{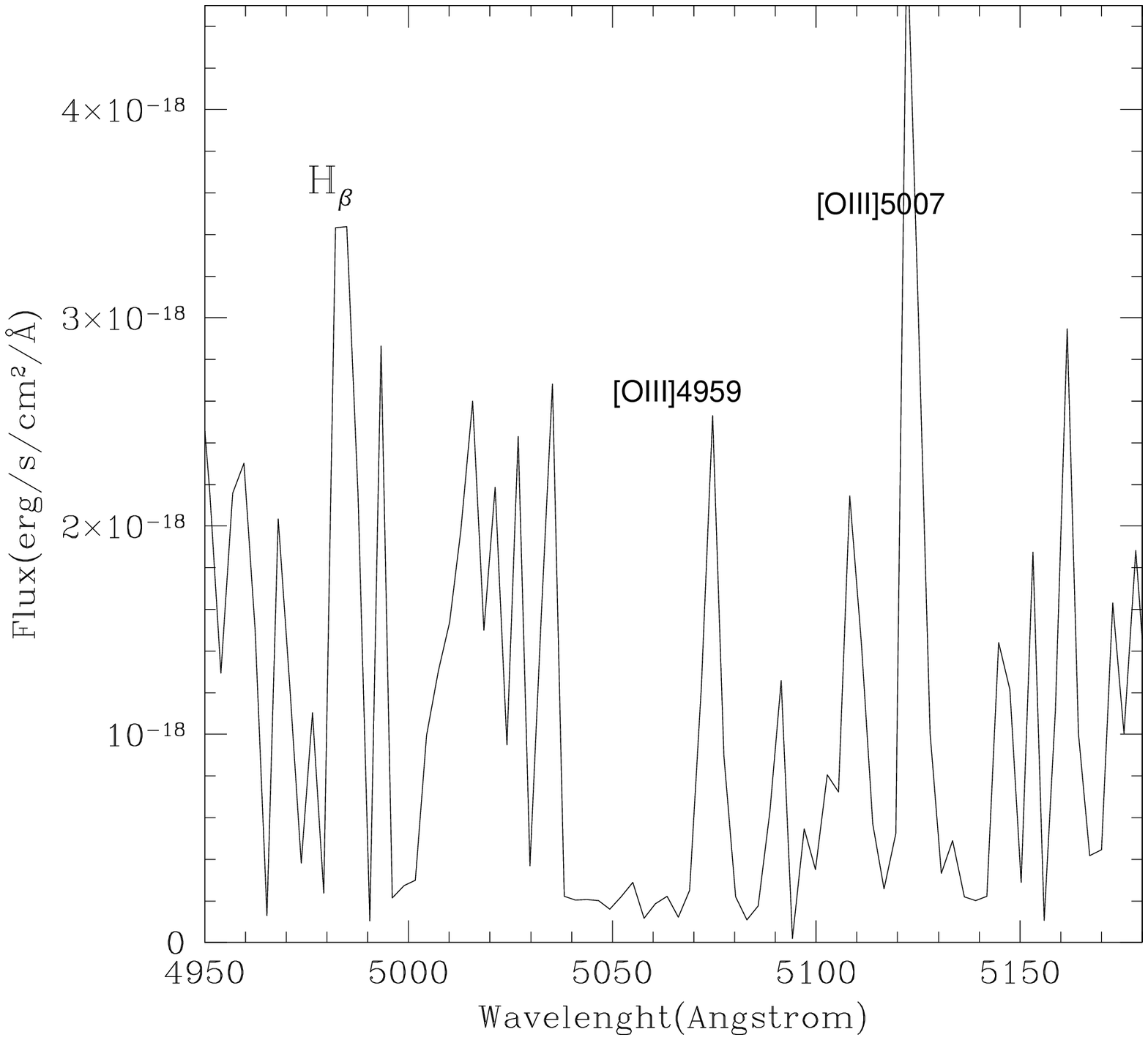}
\caption {Spectrum of UGC7576 obtained by summing up 1D spectra
    extracted along both the North and South side of the polar
    structure.} \label{spec1}
\end{figure*}

\begin{figure*}
\includegraphics[width=8cm]{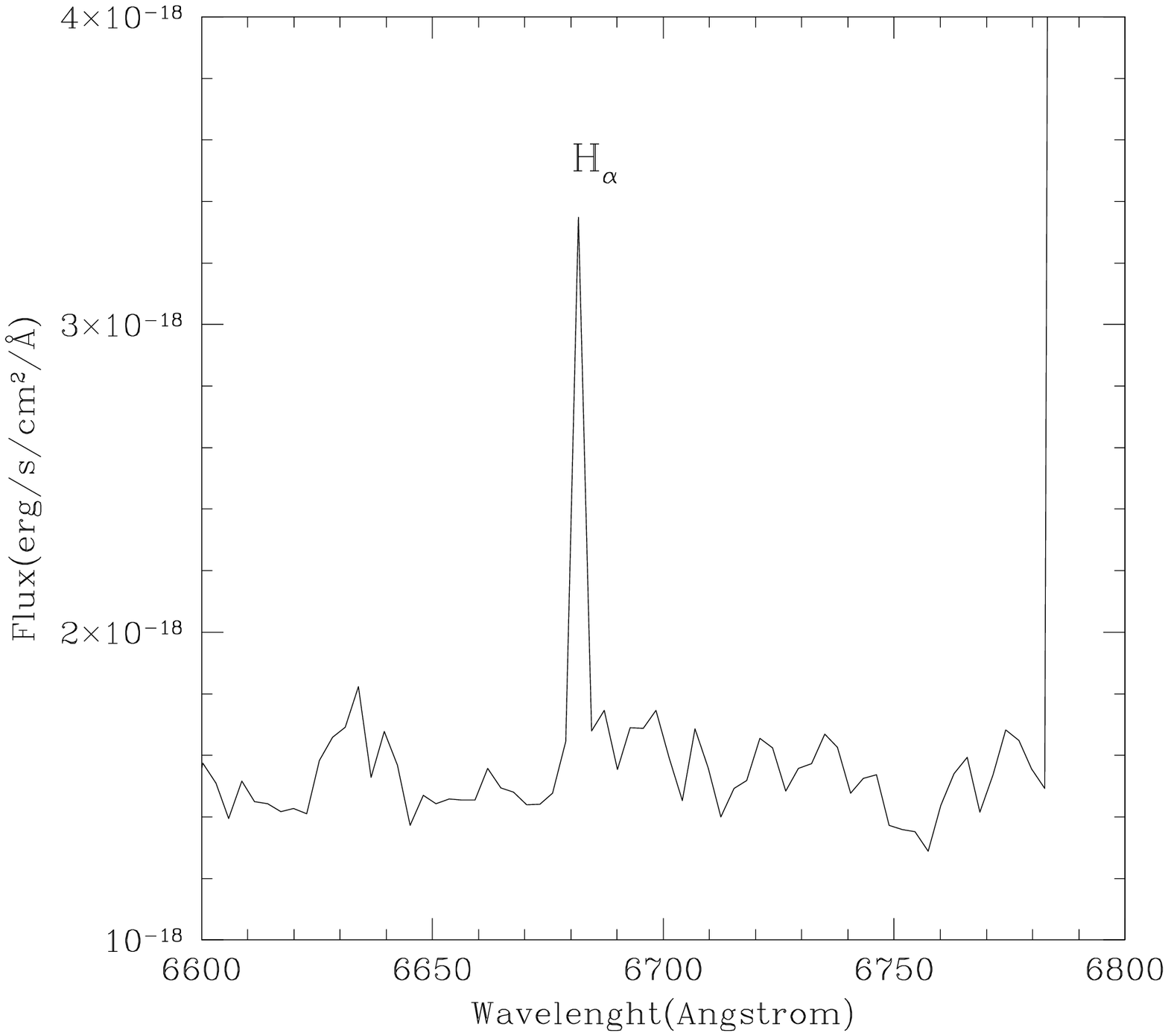}
\includegraphics[width=8cm]{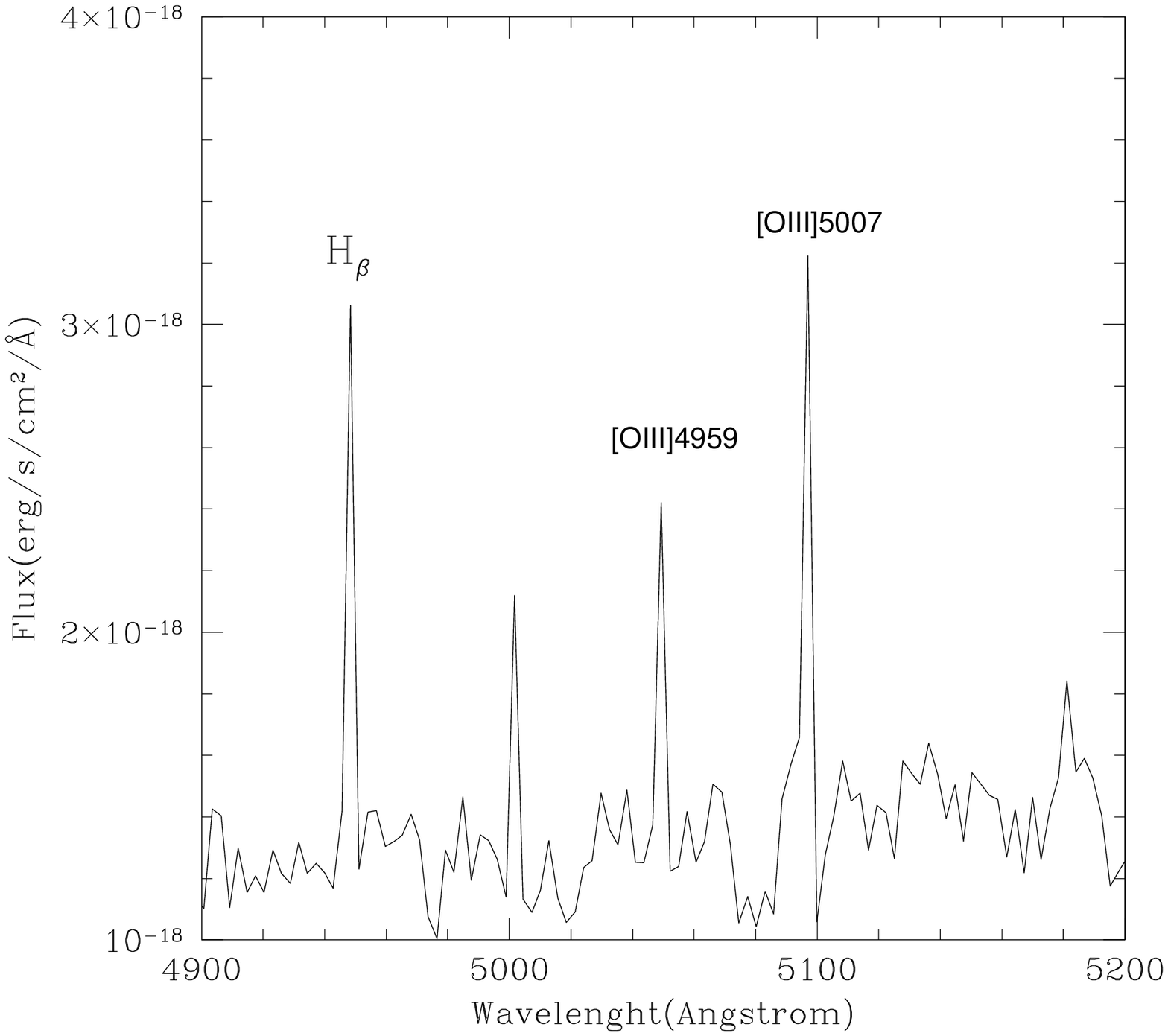}
\caption {Spectrum of UGC9796 obtained by summing up 1D spectra
    extracted along both the North and South side of the polar
    structure.} \label{spec2}
\end{figure*}

The final step of the data-processing is the flux calibration of each
2D spectra, by using observations of the standard star Feige66 and the
standard tasks in IRAF (STANDARD, SENSFUNC and CALIBRATE). To perform
the flux calibration we extracted a 1-D spectrum of the standard star
to find the calibration function; then we extracted a set of 1-D
spectra of the galaxy summing up a number of lines corresponding to
the slit width. Since the slit width was $2 ''$ and the scale of the
instrument was $0.252 ''/pix$, we collapsed eight lines to obtain each
1-D spectrum. Finally we applied the flux calibration to this
collection of spectra. The wavelength and flux-calibrated spectra are shown in
Fig. \ref{spec1} and Fig. \ref{spec2}.

The fluxes of the above mentioned emission lines were measured using
the IRAF {\small SPLOT} routine, that provides an interactive facility
to display and analyze spectra. We evaluated
flux and equivalent width by marking two continuum points around the
line to be measured.  The linear continuum is subtracted and the flux
is determined by simply integrating the line intensity over the local
fitted continuum. The errors on these quantities are calculated,
following \cite{Per03}, by the relation $\sigma_{1} =
\sigma_{c}N^{1/2}[1+EW/(N\Delta)]^{1/2}$, were $\sigma_{1}$ is the
error in the line flux, $\sigma_{c}$ is the standard deviation in a
box near the measured line and represents the error in the continuum
definition, N is the number of pixels used to measure the flux, EW is
the equivalent width of the line and $\Delta$ is the wavelength
dispersion in \AA/pixel.

\subsection{Reddening correction} \label{red}

Reduced and flux calibrated spectra and the measured emission line
intensities were corrected for the reddening, which account both for
that intrinsic to the source and to the Milky Way. By comparing the
intrinsic Balmer decrement $H_{\alpha}/H_{\beta}=2.89$, we derived the
visual extinction $A(V)$ and the color excess $E(B-V)$, by adopting
the mean extinction curve by \cite{Card89} $A(\lambda)/A(V) =
a(x)+b(x)R_{V}$, where $R_{V}[\equiv\ A(V)/E(B-V)]=3.1$ and
$x=1/\lambda$. All the emission lines in our spectra are in the
\emph{optical/NIR} range (see \citealt{Card89}), so we used the average
$R_{V}$-dependent extinction law derived for these intervals to
perform the reddening correction.

We derived the average observed Balmer decrements for both galaxies,
which are the following:\\ $(H_{\alpha}/H_{\beta})_{UGC7576}$ = $3.18
\pm\ 2.10$\\ $(H_{\alpha}/H_{\beta})_{UGC9796}$ = $2.24
\pm\ 1.30$\\ while the color excess obtained by using these observed
decrements are:\\ $[E(B-V)]_{UGC7576}$ = $0.09
\pm\ 0.36$\\ $[E(B-V)]_{UGC9796}$ = $-0.25 \pm\ 0.92$.\\ The
  negative value of the color excess for UGC9796 indicates the
  presence of stars that are bluer and hotter than normal, and thus
  they have V apparent magnitude greater than the B one, leading to a
  negative B-V color. In general, stars hotter than Vega, that have
  $E(B-V)=0$, have negative color excess \citep{Mas98}.

Such values of E(B-V) are used to derive the extinction $A_{\lambda}$,
through the Cardelli's law. Finally, the corrected fluxes are given by

\begin{equation}
\frac{F^{\lambda}_{int}} { F^{H_{\beta}}_{int}} = \frac{F^{\lambda}_{obs}}  {F^{H_{\beta}}_{obs}} 10^{0.4[A_{\lambda}-A_{H_{\beta}}]}
\end{equation}

\section{Empirical oxygen abundances determination}\label{oxy}

The analysis of nebular spectra in HII regions is the best tool for
the determination of chemical abundances in spiral and irregular
galaxies. The abundances of several elements can be determined by
using strong emission lines clearly visible in the spectra.

The main aim of this work is to measure the \emph{Oxygen abundance
  parameter} $R_{23} = ([OII]\lambda 3727 + [OIII]\lambda \lambda 4959
+ 5007)/H_{\beta}$ \citep{Pag79}, and consequently the oxygen
abundance $12+log(O/H)$ and the metallicity of the HII regions in the
polar disks of UGC7576 and UGC9796, following the procedure outlined
by \cite{Spav10} for NGC4650A.

The \emph{Empirical methods} are based on the cooling properties of
ionized nebulae which translate into a relation between emission-line
intensities and oxygen abundance. Several abundance calibrators have
been proposed based on different emission-line ratios: $R_{23}$
\citep{Pag79}, $S_{23}$ \citep{Diaz00}; among the other, in this work
we used the so called P-method introduced by \cite{Pil01}.

\cite{Pil01} realized that for fixed oxygen abundances the value
of $X_{23} = log R_{23}$ varies with the excitation parameter $P =
R_{3}/R_{23}$, where $R_{3} = OIII[4959+5007]/H_{\beta}$, and proposed
that this latter parameter could be used in the oxygen abundance
determination. This method, called ``P-method'', propose to use a more
general relation of the type $O/H = f(P, R_{23})$, compared with the
relation $O/H = f(R_{23})$ used in the $R_{23}$ method. The equation
related to this method is the following
\begin{equation}\label{pil-cal}
    12+log(O/H)_{P} = \frac{R_{23}+54.2+59.45P+7.31P^{2}}{6.07+6.71P+0.371P^{2}+0.243R_{23}}
\end{equation}

where $P = R_{3}/R_{23}$. It can be used for oxygen abundance
determination in moderately high-metallicity HII regions with
undetectable or weak temperature-sensitive line ratios
\cite{Pil01}. The definition of moderately high metallicity is
  adopted from \citet{Pil01} and refers to the abundance interval $8 <
  12+logO/H < 8.5$ where the relation $O/H = F(R_{23})$ \citep{Pag79}
  becomes degenerate. As suggested by \citet{Pil01}, the positions of
  HII regions in the $P-R_{3}$ diagram are related to their oxygen
  abundances. Since both PRGs are in the region where moderately
  high-metallicity HII regions are located, we can use
  Eq. \ref{pil-cal} above to estimate the oxygen abundances for both
  UGC7576 and UGC9796.

We estimate the mean oxygen abundance parameter, $R_{23}$, by
  summing the fluxes of the nebular emission lines at different
  regions along the polar structures (see Tab. \ref{fluxes}). The
average values of oxygen abundance obtained for UGC7576 and UGC9796
are $12+log(O/H)_{P} = 8.5 \pm 0.5$ and $12+log(O/H)_{P} = 7.7 \pm 1$
respectively (see Tab. \ref{ox}). Moreover, for UGC7576 it has been
possible to measure the oxygen abundance $12+log(O/H)_{P}$ at
different distances from the center of the galaxy and the values are
shown in Fig. \ref{fit}.

\begin{table*}[h]
\caption{\label{fluxes}Observed and De-Reddened emission line fluxes relative to $H_{\beta}$.} \centering
\begin{tabular}{lccc}
\hline\hline
line& $\lambda$ (\AA) & Observed flux relative to $H_{\beta}$ & De-Reddened flux relative to $H_{\beta}$\\
\hline
 UGC9796 & & & \\
\hline
$[OII]$ & 3727 & 8.9 &6.8\\
$[OIII]$ & 4959 & 5.0 & 5.1\\
$[OIII]$ & 5007 & 8.7 & 9.0\\
$H_{\gamma}$ & 4340 & 0.9 & 0.8\\
$H_{\alpha}$ & 6563 & 2.2 & 2.8\\
\hline
 UGC7576 & & & \\
\hline
$[OII]$ & 3727 & 2.5 & 2.8\\
$[OIII]$ & 4959 & 1.5 & 1.51\\
$[OIII]$ & 5007 & 1.7 & 1.71\\
$H_{\gamma}$ & 4340 & 1.1 & 1.15\\
$H_{\alpha}$ & 6563 & 3.2 & 2.9\\
\hline
\end{tabular}
\end{table*}

\begin{table*}[h]
\caption{\label{ox}Oxygen abundances and metallicities of UGC7576 and UGC9796.} \centering
\begin{tabular}{lcc}
\hline\hline
Parameter&UGC7576&UGC9796\\
\hline
$12+log(O/H)$\tablefootmark{a} & $8.5 \pm 0.5 $ & $7.7 \pm 1 $\\
$Z/Z_{\odot}$\tablefootmark{a} & $0.4 \pm 0.002 $ & $0.1 \pm 0.001 $\\
\hline
\end{tabular}
\tablefoot{\tablefoottext{a}{Spavone et al. (2010) derived the following values of oxygen abundance and metallicity for NGC4650A: $12+log(O/H) = 8.2 \pm 0.2$ and $Z/Z_{\odot} = 0.2 \pm 0.002$}}
\end{table*}


\section{Metallicity and SFR estimates}\label{disc}

We derive the
oxygen abundance $12+log(O/H)$ for UGC7576 and UGC9796, by using the
empirical P-method \citep{Pil01}. We found $12+log(O/H) = 8.5 \pm 0.5$
for UGC7576 and $12+log(O/H) = 7.7 \pm 1$ for UGC9796. The
metallicities corresponding to each value of oxygen abundances given
before have been estimated. We adopted $12 + log(O/H)_{\odot} = 8.83 \pm 0.20=
A_{\odot}$ and $Z_{\odot} = 0.02$ \citep{Asp04}. Given that $Z \approx
K Z_{\odot}$, $K_{UGC7576} = 10^{[A_{UGC7576} - A_{\odot}]}$ and
$K_{UGC7576} = 10^{[A_{UGC9796} - A_{\odot}]}$, we obtain
metallicities for the HII regions of the polar disk of $Z \simeq
0.008$ in UGC7576 and $Z \simeq 0.002$ in UGC9796 which correspond
respectively to $Z \simeq (0.400 \pm 0.002) Z_{\odot}$ and $Z \simeq
(0.100 \pm 0.001) Z_{\odot}$, in good agreement with the values obtained
by \cite{Rad03} (see Tab. \ref{ox}).

\subsection{Metallicity-luminosity relation}
\begin{figure*}
\centering
\includegraphics[width=12cm]{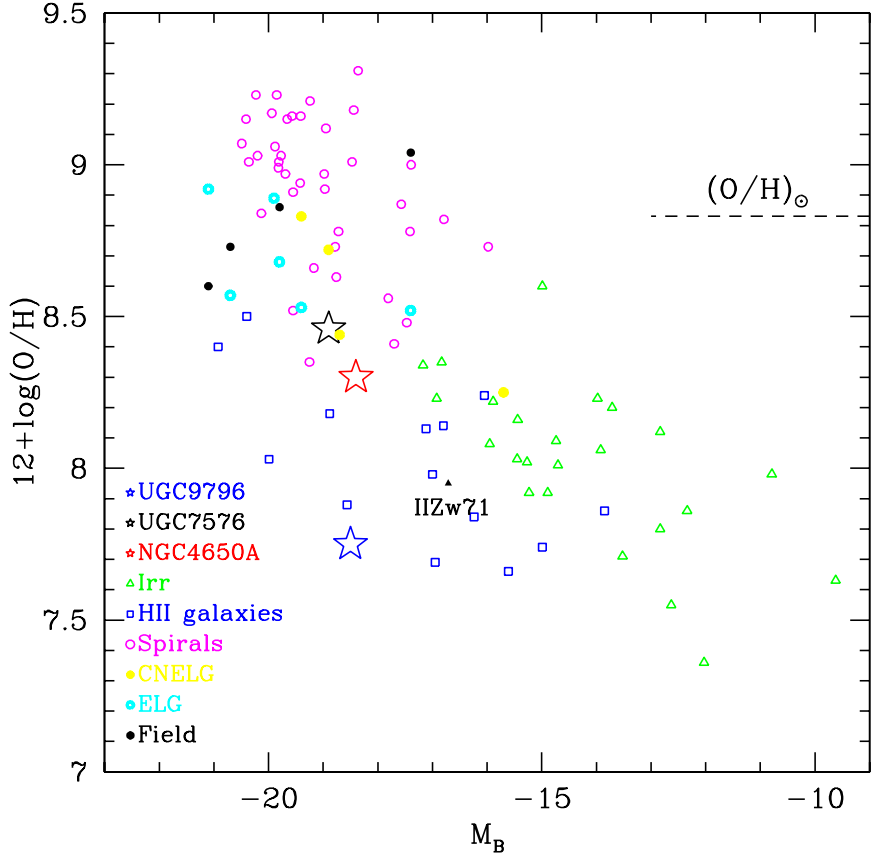}
\caption{Oxygen abundance vs absolute blue magnitude for Compact
  Narrow Emission-Line Galaxies (CNELGs, yellow filled circles),
  star-forming Emission Line Galaxies (ELGs, cyan open circles), four
  field galaxies with emission lines (filled black circles), nearby
  dwarf irregulars (open triangles), local spiral galaxies (open
  circles), local HII galaxies (open squares), NGC4650A (red star)
  (Spavone et al. 2010), the polar disk galaxy IIZw71 (Perez-Montero
  et al.2009), UGC7576 (black star) and UGC9796 (blue star) (this
  work). The dashed line indicates the solar oxygen abundance.} \label{conf}
\end{figure*}

The mean values for the oxygen abundance along the polar structure of
UGC7576 and UGC9796, derived by the empirical method (see
sec. \ref{oxy}), are compared with those for a sample of late-type
disk galaxies by \cite{Kob99}, as function of the total
luminosity. To this aim, we converted the absolute blue magnitude
  for the objects in the sample of \citet{Kob99} by using $H_{0} =75$
  km/s/Mpc. Results are shown in Fig. \ref{conf}.

We found that UGC7576 is located in the region where spiral galaxies
are found, and, contrary to its high luminosity, it has metallicity
lower than spiral galaxy disks of the same total luminosity and it is
consistent with that observed for NGC4650A.

UGC9796 instead is even more metal-poor than UGC7576 and NGC4650A, in
fact it is located in the region where HII and irregular galaxies are
also found, characterized by lower luminosities and metallicities with
respect to the spiral galaxies.

We also compare our new results with those obtained by \cite{Per09} for
IIZw71, a blue compact dwarf galaxy also catalogued as a probable
polar ring: consistently with its
low luminosity, the metallicity of the brightest knots in the ring of
IIZw71 is lower with respect to that of UGC7576, but it is slightly
higher than those of UGC9796. Taking into account the total magnitude,
such values are somewhat lower than that expected by the
metallicity-luminosity relation.

\subsection{Star formation rate}

The $H_{\alpha}$ emission is detected in both systems with adequate
signal-to-noise, and from the measured integrated flux we can derive
the star formation rate in the polar ring structure. We have derived
the SFR for the polar structures of UGC7576 and UGC9796, from the
$H_{\alpha}$ luminosity using the expression given by \cite{Ken98}
$SFR = 7.9 \times 10^{-42} \times L(H_{\alpha})$. We find that it is
almost constant along the disks of both galaxies, within a large
scatter in the individual values. From the average values of
$L(H_{\alpha}) \simeq 2.36 \times 10^{36}$ erg/s for UGC7576 and
$L(H_{\alpha}) \simeq 5.33 \times 10^{35}$ erg/s for UGC9796, we have
obtained an average $SFR \sim 1.9 \times 10^{-5} M_{\odot}/yr$ and
$SFR \sim 4.2 \times 10^{-6} M_{\odot}/yr$ respectively. These values
are significantly lower than those obtained for NGC4650A
\citep{Spav10}, which is $SFR \sim 0.06 M_{\odot}/yr$ and IIZw71
\citep{Per09}, which is $SFR \sim 0.035 M_{\odot}/yr$.

Taking into account that the polar structure in both PRGs is
very young, since the last burst of star formation occurred less than
1 Gyr ago \citep{Res94}, we check if the present SFR and even 2 and 3
times higher (i.e. $SFR = 3.8 \times 10^{-5} M_{\odot}/yr$, $SFR = 5.7
\times 10^{-5} M_{\odot}/yr$ and $SFR = 8.4 \times 10^{-6}
M_{\odot}/yr$, $SFR = 1.26 \times 10^{-5} M_{\odot}/yr$), can give the
inferred metallicities of $Z=0.4 Z_{\odot}$ (for UGC7576) and $Z=0.1
Z_{\odot}$ (for UGC9796) and how strongly could increase the
metallicity with time.

Given the presence of star forming regions in the polar disk of
NGC4650A, \cite{Spav10} used a constant SFR law for this object. In
the case of UGC7576 and UGC9796, the redder B-V color and the
  lower $M_{HI}/L_{B}$ suggest a longer time since the last burst of
  star formation in there objects. Furthermore the difference in the
  detected $H_{\alpha}$ fluxes and consequently of the SFR of these
  galaxies are consistent with it (see Tab. \ref{global}).  Given
  these observational properties and the absence of star forming
  clumps, we adopted an exponentially declining SFR started several
  Gyrs ago: this is typically used for late-type galaxies because, on
  average, it gives even older stars than a linear decay. The
  expression is: $SFR(t) = M_{\star}\tau^{-1}exp[-(t_{0}-t)/\tau)]$ ,
  where we assume $t_{0} = 8 Gyrs$, because in $\Lambda$CDM models
  such cold accretion is more unlikely at low redshifts, $\tau =
  2Gyrs$ is the decay timescale and $t$ is the lookback time
  \citep{Bru03}.

 By using the mass-metallicity relation derived
  by \cite{Tre04}, where $12+log(O/H) = -1.492 + 1.847 log(M_{\star})
  - 0.08026 (log M_{\star})^{2}$, we found that for UGC7576 $0.5
  Z_{\odot} \le Z \le 1 Z_{\odot}$ and for UGC9796 $0.05 Z_{\odot} \le
  Z \le 0.45 Z_{\odot}$. This shows that the present SFR for the polar
  structures is able to increase the metallicity of about 0.15
  $Z_{\odot}$ for UGC7576 and 0.05 $Z_{\odot}$ for UGC9796, after
  1Gyr. We also derived the expected metallicities by using different
  values for the free parameters $\tau$ (the decay timescale) and
  $t_{0}$, in order to check how these parameters can change the
  metallicity range: we found that the obtained metallicities are
  comparable, or even lower, than those reported above, with
  differences of about 4\%.

For UGC7576 the derived values for Z are larger than {\bf $Z=0.400 \pm
  0.002 Z_{\odot}$} found by using the element abundances (see
  Sec.\ref{oxy}); for UGC9796, instead, we found that the metallicity
  of {\bf $Z=0.100 \pm 0.001 Z_{\odot}$}, estimated by using the element
  abundances, falls near the lower limit of the range of expected
  metallicities. The implications of these results will be discussed
  in details in Sec. \ref{conc}.

\subsection{Metallicity gradient along the polar structure of UGC7576}
\begin{figure*}
\centering \includegraphics[width=10cm]{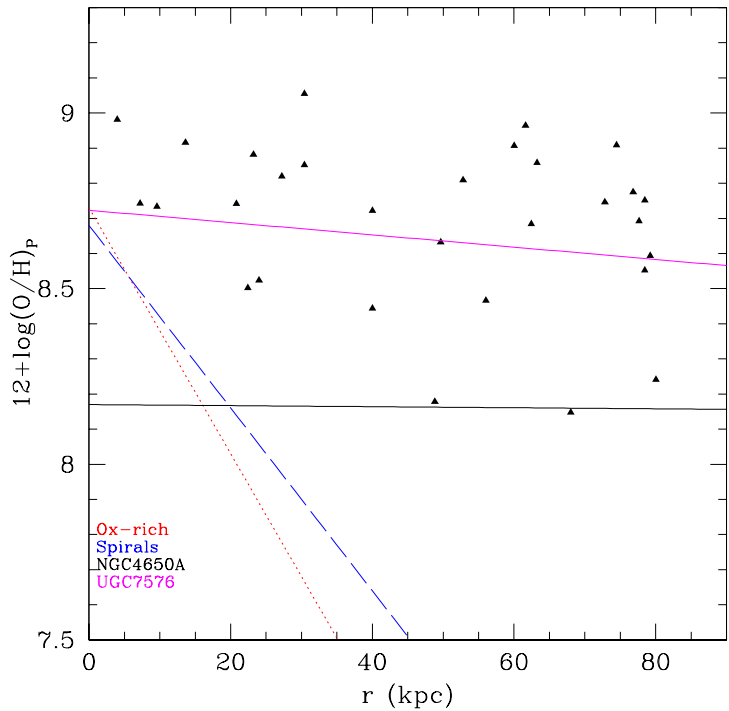}
\caption{Oxygen abundance of UGC7576 derived with P- method, proposed
  by Pilyugin (2001), versus radius. The superimposed lines are the
  linear best fit derived by Pilyugin et al. (2006); the red line
  represents the best fit to the abundance of oxygen-rich spirals,
  while the blue line is those related to ordinary spirals. The black
  line is the best fit obtained for NGC4650A, while the magenta line
  is the best fit obtained for UGC7576.} \label{fit}
\end{figure*}

The oxygen abundance in the polar ring of UGC7576 as a function of the
radius derived by empirical methods is shown in Fig. \ref{fit}: the
plot shows that the metallicity remains almost constant along the
  projected major axis of the polar ring. It is only slightly
declining, differently from what it is observed for spiral galaxies,
where there is a very steep gradient in metallicity. The absence of
metallicity gradient is a typical behavior found in LSB galaxies
\citep{deB98}, and also in the polar disk galaxy studied by
\cite{Bro09} and in NGC4650A \citep{Spav10}. This suggests that, as
already pointed out for NGC4650A \citep{Spav10}, the star formation
and the metal enrichment in the polar structure is not influenced by
the stellar evolution of the central spheroid and that the polar ring
was formed later. As also suggested by \cite{Rup10}, numerical
simulations predict that the absence of metallicity gradient is
accounted for by inflow of low metallicity gas from the outskirts.

On the contrary, ordinary and oxygen-rich spiral galaxies, where
  no primordial gas is accreted from outside, show a decreasing
  abundance with increasing radii (see Fig. \ref{fit} and
  \citealt{Pil06}). These observed features in spiral disks are well
  explained by the infall models of galaxy formation: they predict
  that these systems build up through the accretion of on-site gas,
  which become more metal rich while it flows towards the galaxy
  center (\citealt{Mat89}; \citealt{Boi99}). Such process generates
  the observed gradients.

\section{Formation history for UGC7576 and UGC9796: discussion and conclusions}\label{conc}
Galaxies with a polar ring/disk have a special role in the studies of
physical processes at work during galaxy interactions and merging. In
order to account both for the featureless morphology of the central
spheroidal galaxy and for the more complex structure of the polar
ring/disk, so far three main formation processes have been
proposed: {\it i)} a major dissipative merger; {\it ii)} tidal
accretion of material (gas and/or stars) by outside; {\it iii)} cold
accretion of pristine gas along a filament.

In the merging scenario, the PRG results from a ``polar'' merger of
two disk galaxies with unequal mass, (\citealt{Bek97};
\citealt{Bek98}; \citealt{Bou05}). In the accretion scenario, the
polar ring/disk may form by a) the disruption of a dwarf companion
galaxy orbitating around an early-type system, or by b) the tidal
accretion of gas stripping from a disk galaxy outskirts, captured by
an early-type galaxy on a parabolic encounter (\citealt{Res97};
\citealt{Bou03}; \citealt{Han09}). Both major merger and accretion
scenarios are able to account for many observed PRGs morphologies and
kinematics, such as the existence of both wide and narrow rings,
helical rings and double rings \citep{Whi90}.

The cold accretion scenario has been proposed very recently for the
formation of a wide disk-like polar rings (see Sec. \ref{intro}): a
long-lived polar structure may form through cold gas accretion along a
filament, extended for $\sim 1$ Mpc, into the virialized dark matter
halo \citep{Mac06}. In this formation scenario, there is no limits to
the mass of the accreted material, thus a very massive polar disk may
develop either around a stellar disk or a spheroid.

As suggested by the previous studies on PRGs (\citealt{Spav10},
\citealt{Iod06}), the critical physical parameters that allow to
discriminate among the three formation scenarios are 1) the total
baryonic mass (stars plus gas) observed in the polar structure with
respect to that in the central spheroid; 2) the kinematics along both
the equatorial and meridian planes; 3) the metallicity and SFR in the
polar structure.\\ In the tidal accretion scenario the total amount of
accreted gas by the early-type object is about $10\%$ of the gas in
the disk donor galaxy, i.e. up to $10^{9}M_{\odot}$. In the case of
UGC7576 and UGC9796, the polar structure is characterized by an high
baryonic mass (see Tab. \ref{scenarios}), comparable with the total
luminous mass in the central spheroid. Similarly to the case of
NGC4650A \citep{Spav10}, the large HI mass exclude the formation of
the polar structures through the tidal accretion of gas and stars by
an external donor galaxy \citep{Bou03}.


In the merging scenario, the morphology and kinematics of the merger
remnants depends on the merging initial orbital parameters and the
initial mass ratio of the two galaxies \citep{Bou05}. In the case
  of UGC7576 and UGC9796, this scenario is ruled out because,
  according to simulations (e.g. \citealt{Bou05}), a high mass ratio
  of the two merging galaxies is required to form a massive and
  extended polar disk as observed in both the PRGs: this would convert
  the intruder into an elliptical-like, not rotationally supported,
  stellar system. Since this is in contrast with the high
maximum rotation velocities observed (see \citealt{Res94} and
Tab. \ref{scenarios}) in UGC7576 ($\sim\ 212 km/s$) and UGC9796
($\sim\ 157 km/s$), the merging scenario is ruled out for both
galaxies.

As already mentioned, if the polar structure, both around an
elliptical and disk galaxy, forms by the cold accretion of gas from
filaments there is no limit to the accreted mass. Moreover, due to the
inflow of pristine gas, the metallicity is lower with respect to that
observed in galaxies of comparable total luminosity, and its value
derived by the present SFR is higher than those directly measured by
the chemical abundances. \cite{Spav10} have found that such
predictions were consistent with observations in the polar disk
galaxy NGC4650A, yielding to the conclusion that the cold accretion of
gas by cosmic web filaments is the most realistic formation scenario
for this object.

By studying the chemical abundances in the polar structure of
UGC7576 and UGC9796, in the present work we aim at testing the cold
accretion scenario for these objects. The main results are as
follows (see also Sec. \ref{disc}): 1) both PRGs have on average
lower metallicity with respect to that of same-luminosity spiral
disks, in particular, for UGC7576 $Z=0.4 Z_{\odot}$ and for
UGC9796 $Z=0.1 Z_{\odot}$; 2) the metallicity expected for the
present SFR at three different epochs are higher than those
measured from the element abundances, and they varies in the range
$0.5 Z_{\odot} \leq Z \le 1 Z_{\odot}$ for UGC7576 and $0.05
Z_{\odot} \le Z \leq 0.45 Z_{\odot}$ for UGC9796; 3) for UGC7576,
the metallicity remains almost constant along the polar structure.

In the following, we will address how these results reconcile with
the theoretical predictions for the cold accretion process and
discuss how the SFR and metallicity derived in this work, together
with the other two key parameters, i.e. the baryonic mass and
kinematics, available by previous studies on these objects, may
help to discriminate among the three formation scenarios, in non
ambiguous way as done for NGC4650A. All these quantities, for both
the PRGs studied in this work and for NGC4650A, are summarized in
Tab. \ref{scenarios}.

The cold accretion mechanism for disk formation predicts rather low
metallicity ($Z=0.1Z_{\odot}$) (\citealt{Dek06}, \citealt{Ocv08},
\citealt{Age09}): such value refers to the time just after the
accretion of a misaligned material, so it can be considered as initial
value for Z before the subsequent enrichment. How this may reconcile
with the observed metallicity for UGC7576 and UGC9796?

We estimated that the present SFR for the polar structure is $SFR =
1.9 \times 10^{-5} M_{\odot} yr^{-1}$ for UGC7576 and $SFR = 4.2
\times 10^{-6} M_{\odot} yr^{-1}$ for UGC9796: these values are able
to increase the metallicity of about $0.15 Z_{\odot}$ for UGC7576 and
$0.05 Z_{\odot}$ for UGC9796 after 1Gyr (see Sec.\ref{disc}): taking
into account that the polar structure is very young, less than 1Gyr
\citep{Res94p}, an initial value of $Z=0.1Z_{\odot}$, at the time of
polar structure formation, could be consistent with the today's
observed metallicity. Therefore, the cold accretion of gas can be a
possible formation scenario for both the PRGs UGC7576 and
UGC9796. In the case of UGC7576, one more hint for the cold
  accretion scenario comes from the fact that the metallicity expected
  by the present SFR turns to be higher than those directly measured
  by the chemical abundances: as already suggested for NGC4650A
  \citep{Spav10}, a possible explanation for this observed feature
  could be the infall of pristine gas (\citealt{Fin07};
  \citealt{Ell08}).\\ Moreover, in the case of UGC7576, the lack of
  abundance gradient along the polar structure is consistent with the
  prediction of recent numerical simulations by \cite{Rup10} and
  \cite{Kew08}; they observe flatter metallicity gradients than those
  observed in typical spiral galaxies, due to the radial inflow of low
  metallicity gas from outside. For UGC9796, instead, the metallicity
  derived by the chemical abundances falls near the lower limit of the
  expected metallicity range; for this reason we cannot definitely
  rule out other possible formation scenarios.

The high baryonic (gas+stars) masses, the large extentions of the
polar structure and the low metallicity observed in these PRGs are not
accounted for in the formation of polar rings through the disruption
of a dwarf galaxy, which are characterized by higher metallicities
($1/9 Z_{\odot} \div\ 1/3 Z_{\odot}$) and lower HI masses ($\sim\
10^{7} M_{\odot}$; \citealt{Gal09} and \citealt{Bek08}).

\begin{figure*}
\centering
\includegraphics[width=10cm]{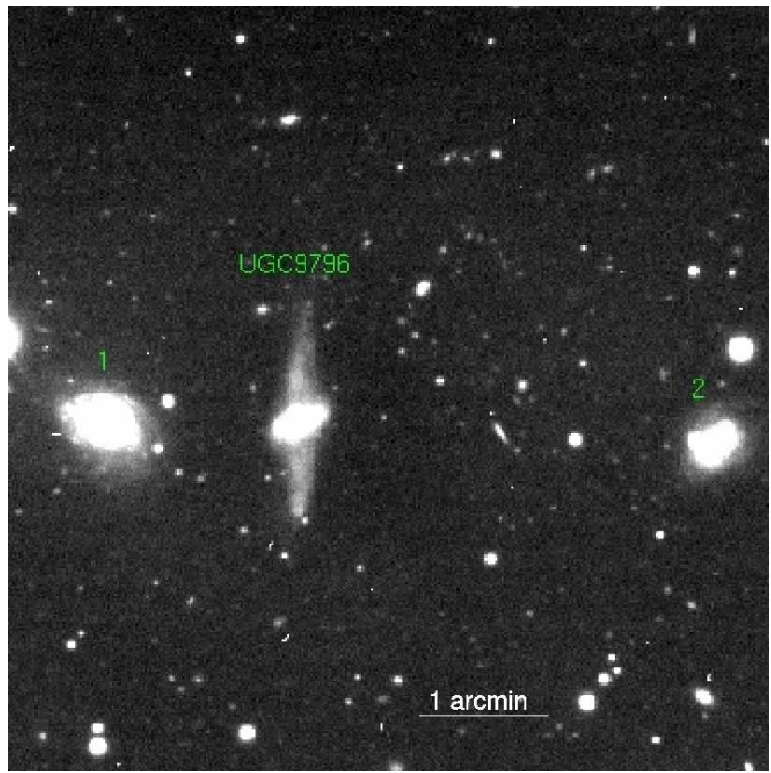}
\caption{Optical image of UGC9796 and its two nearest companion
  galaxies, MCG +07-31-049 (number 1) and CGPG 1514.2+4320 (number
  2).} \label{comp}
\end{figure*}

Differently, the tidal accretion scenario, in which gas is stripped
from a gas-rich donor in a particular orbital configuration
\citep{Bou03}, is able to produce wide polar rings and/or disks both
around a disk or an elliptical galaxy.  In order to consider the
accretion hypothesis, we studied the field around these galaxies to
see if there are any objects as possible donor galaxy candidates:
inside a radius of about 5 times the diameter of both PRGs, as
suggested by \cite{Broc97}, we find that UGC7576 has no close
companions, while UGC9796, that is in a small group, has 5 companions
\citep{Cox06}. Therefore, while also the tidal accretion scenario is
ruled out for UGC7576, it needs to be envisioned in the case of
UGC9796.\\
The VLA images show that UGC9796 and the close companion galaxy
MCG+07-31-049 (labelled as 1 in Fig. \ref{comp}) could be in the
orbital configuration needed to form a polar structure through an
accretion event \citep{Bou03}: in fact, the HI gas in
UGC9796, which is all associated with the polar structure, has the
outer contours warped away from the poles and, in the SE regions,
it extends towards MCG+07-31-049, which has HI distribution
extending in a direction perpendicular to that of UGC9796.

In the tidal accretion process, the accreted gas comes only from
the outer and more metal-poor regions of the donor galaxy. All the
companion galaxies of UGC9796 have an amount of HI gas ($\sim\ 10^{9}
M_{\odot}$) comparable with those of this PRG \citep{Cox06}, but,
according to \cite{Bre09}, the metallicity of the outer regions of
bright spiral galaxies is $0.2 Z_{\odot} \leq\ Z \leq\ 1.1 Z_{\odot}$
and the observed value for UGC9796, {\bf $Z = 0.100 \pm 0.001 Z_{\odot}$}, is
below the lower limit.

Given all the evidences shown above, we can give the following
conclusions for this work.\\ The cold accretion of gas by cosmic web
filaments could well account for both the low metallicity, the lack of
gradient and the high HI content in UGC7576. Moreover, the general
underdensity of the environment where this galaxy is, can be
consistent with the cold flow accretion of gas as the possible
formation mechanism for this object.\\ For UGC9796 instead, the
scenario is slightly more complex. In fact, the low metallicity
estimated, even lower than those observed in the outskirts of spiral
galaxies is consistent with the formation of disks through cold
accretion mechanisms (\citealt{Dek06}, \citealt{Ocv08} and
\citealt{Age09}). On the other hand, the tidal accretion scenario
cannot be ruled out, given that the galaxy MCG +07-31-049, with its
high amount of HI and its orbital configuration, could be a good
candidate donor for UGC9796.

\begin{table*}
\caption{Discriminant parameters between different formation scenarios}   
\label{scenarios}      
\centering                          
\begin{tabular}{c c c c c c c c c c}        
\hline\hline                 
PRG & $M_{b}^{HG}$ & $M_{b}^{PD}$ & $V_{eq}$ & $V_{eq}/V_{PD}$ & $\sigma_{0}$ & $M_{b}^{HG}/M_{b}^{PD}$ & $v/ \sigma$ & $Z_{est}$ & $Z_{exp}$ \\    
 & $(M_{\odot})$ & $(M_{\odot})$ & (km/s)& & (km/s)& & &$(Z_{\odot})$ & $(Z_{\odot})$\\
\hline                        
   UGC7576 & $7.86\times 10^{9}$ & $2.88\times 10^{9}$ & 212 & 0.96 & 116 & 2.73 & 1.8 & $(0.400\pm 0.002)$ & $(0.5\div 1)$ \\      
   UGC9796 &$1.0\times 10^{10}$ & $3.05\times 10^{9}$ & 157 & 1.08 & 73 & 3.28 & 2.15 & $(0.100\pm 0.001)$ & $(0.05\div 0.45)$ \\
   NGC4650A &$5\times 10^{9}$ &$12\times 10^{9}$ &90 &0.75 & 70 &0.42&1.28 & $(0.200\pm 0.002)$ & $(1.02\div 1.4)$  \\
\hline                                   
\end{tabular}
\end{table*}

\begin{acknowledgements}
The authors wish to thank the anonymous referee for the detailed and
constructive report, and Frederic Bournaud, for many useful discussions and
suggestions. M. S. and E. I. wish to thank Massimo Capaccioli for his
useful suggestions, which allowed to improve this paper. This work is
based on observations made with Telescopio Nazionale Galileo (TNG)
under programme ID A21TAC-54.
\end{acknowledgements}


\end{document}